\documentclass[prc,amsmath,showkeys,showpacs,superscriptaddress,floatfix,letterpaper]{revtex4}

\usepackage{dcolumn}
\usepackage{graphicx}

\begin {document}
  \newcommand {\nc} {\newcommand}
  \nc {\beq} {\begin{eqnarray}}
  \nc {\eeq} {\nonumber \end{eqnarray}}
  \nc {\eeqn}[1] {\label {#1} \end{eqnarray}}
  \nc {\eol} {\nonumber \\}
  \nc {\eoln}[1] {\label {#1} \\}
  \nc {\ve} [1] {\mbox{\boldmath $#1$}}
  \nc {\mrm} [1] {\mathrm{#1}}
  \nc {\half} {\mbox{$\frac{1}{2}$}}
  \nc {\thal} {\mbox{$\frac{3}{2}$}}
  \nc {\fial} {\mbox{$\frac{5}{2}$}}
  \nc {\la} {\mbox{$\langle$}}
  \nc {\ra} {\mbox{$\rangle$}}
  \nc {\etal} {\emph{et al.\ }}
  \nc {\eq} [1] {(\ref{#1})}
  \nc {\Eq} [1] {Eq.~(\ref{#1})}
  \nc {\Ref} [1] {Ref.~\cite{#1}}
  \nc {\Refc} [2] {Refs.~\cite[#1]{#2}}
  \nc {\Sec} [1] {Sec.~\ref{#1}}
  \nc {\chap} [1] {Chapter~\ref{#1}}
  \nc {\anx} [1] {Appendix~\ref{#1}}
  \nc {\tbl} [1] {Table~\ref{#1}}
  \nc {\fig} [1] {Fig.~\ref{#1}}

  \nc {\bfig} {\begin{figure}}
  \nc {\efig} {\end{figure}}
  \nc {\ex} [1] {$^{#1}$}
  \nc {\Sch} {Schr\"odinger }
  \nc {\flim} [2] {\mathop{\longrightarrow}\limits_{{#1}\rightarrow{#2}}}
\title{Peripherality of breakup reactions}
\author{P.~Capel \footnote{Present address:
Physique Quantique, C.P. 165/82 and
Physique Nucl\'eaire Th\'eorique et Physique Math\'ematique, C.P. 229,
Universit\'e Libre de Bruxelles (U.L.B.), B-1050 Brussels, Belgium}}
\email{pierre.capel@centraliens.net}
\affiliation{TRIUMF, 4004 Wesbrook Mall, Vancouver, B.C., Canada V6T2A3}
\author{F.~M.~Nunes}
\email{nunes@nscl.msu.edu}
\affiliation{National Superconducting Cyclotron Laboratory
and Department of Physics and Astronomy,
Michigan State University, East Lansing, Michigan 48824, USA}
\date{\today}
\begin{abstract}
The sensitivity of elastic breakup to the interior of the
projectile wave function is analyzed.
Breakup calculations of loosely bound nuclei ($^8$B and $^{11}$Be)
are performed with two different descriptions of the projectile.
The descriptions differ strongly in the interior of the wave function,
but exhibit identical asymptotic properties, namely the same
asymptotic normalisation coefficient, and phase shifts.
Breakup calculations are performed at intermediate energies
(40--70~MeV/nucleon) on lead and carbon targets as well
as at low energy (26~MeV) on a nickel target. No dependence on the
projectile description is observed. This result confirms that
breakup reactions are peripheral in the sense that they
probe only the external part of the wave function.
These measurements are thus not directly sensitive to the total
normalization of the wave function, i.e. spectroscopic factor.
\end{abstract}
\maketitle

\section{Introduction}
The development of radioactive ion beams in the mid-80s
has allowed the study of nuclei far from stability.
This technical breakthrough led to the discovery
of halo nuclei on the neutron-rich side of the valley
of stability \cite{tan85b,tan85r}.
These loosely bound nuclei have a strongly clusterized
structure \cite{HJJ95,Tan96,Jon04,BHT03}.
In a simple model, they are seen as a core, that contains most
of the nucleons, to which one or two neutrons are loosely bound.
Due to this small binding, the valence neutrons
tunnel far outside the classically allowed region and form
a sort of halo around the core \cite{HJ87}.
Although less probable, proton halos are also possible.

Being very short lived, these nuclei can be studied
only by indirect techniques like elastic breakup.
In this reaction, the halo dissociates from the core
through interaction with a target \cite{HJJ95,Tan96,Jon04,BHT03}.
From the comparison of the data with the theoretical prediction,
one usually extracts a spectroscopic factor for the single-particle
configuration of the projectile model \cite{Nak94,Pal02,Pra03,Fuk04}.
Alternatively, it has been suggested that breakup reactions
being highly peripheral,
only the asymptotic normalization
coefficient (ANC) can be extracted from the measurements
\cite{TCG01,TCG04}. These two viewpoints are contradictory
and must be disentangled.

Recent theoretical papers show that breakup calculations
are sensitive mainly to asymptotic properties
of the projectile \cite{TB04,TB05,CN06}.
Both the ANC of the initial bound state, and the phase shifts
in the continuum play significant roles in the dissociation.
However, they do not prove the breakup not
to be sensitive to the interior of the bound-state wave function.

In most reaction studies, peripherality of a reaction refers to insignificant
contributions to the breakup cross section from small distances
between the projectile and the target (e.g. \cite{Bau84,SL98}).
However here we are interested in peripherality
relative to the internal coordinate of the projectile, the distance between
the core and the valence nucleon. This notion of peripherality is intimately
related to the sensitivity of the reaction to the interior of the
projectile wave function.
Since it is not directly related to the projectile-target peripherality,
it deserves particular attention.

To analyze the role played by the
internal part of the wave function in the dissociation,
we perform breakup calculations using two
descriptions of the projectile that differ only
in the interior of the wave function.
To obtain such a pair of descriptions, we use
the supersymmetric transformations developed by Baye \cite{Bay87l,Bay87a}.
Their great advantage
is that the initial Hamiltonian and its supersymmetric partner
exhibit identical asymptotic properties.
Therefore, both descriptions exhibit the same phase shifts and ANC,
but differ strongly in the interior.

The difference between calculations performed using both descriptions
will tell us about the sensitivity to the interior of
the projectile wave function.
In this study, we consider two projectiles:
a proton-halo nucleus (\ex{8}B),
and a neutron-halo nucleus (\ex{11}Be).
The calculations are performed at intermediate (40--70~MeV/nucleon)
and low (26~MeV) energies using either the dynamical eikonal model
of breakup \cite{BCG05,GBC06} or the Continuum Discretized
Coupled Channel technique (CDCC) \cite{Kam86,TNT01}.
We also consider Coulomb and nuclear dominated reactions.
In each case, the differences between both descriptions are
analyzed through different observables (energy, angular,
and parallel-momentum distributions).

In \Sec{P}, we present the projectile descriptions
used in the calculations, including a summary on the supersymmetric
transformations.
The results of our analysis are displayed in \Sec{results}.
The conclusions are drawn in the last section.

\section{Projectile descriptions}\label{P}

\subsection{Supersymmetric transformations}\label{susy}
In both the dynamical eikonal \cite{BCG05,GBC06},
and the CDCC \cite{Kam86,TNT01} models of
breakup reactions, the projectile is described
as a two-body system: a spherical, structureless core $c$
to which a pointlike fragment $f$ is loosely bound.
The internal structure of the projectile is modeled by
the Hamiltonian
\beq
H_0=-\frac{\hbar^2}{2\mu}\Delta+V_0(\ve{r}),
\eeqn{1}
where $\ve{r}$ is the relative coordinate of the fragment to
the core, and $\mu$ is the reduced mass
of the $c$-$f$ system.
The potential $V_0$ describes the interaction between
the core and the fragment.
It is adjusted to reproduce the loosely bound states,
and some of the resonances of the projectile.
It contains both Coulomb and nuclear terms.
For the present study, we choose $V_0$ deep enough
to obtain an additional deeply bound state in the
partial wave of the physical ground state.
This spurious state has no physical meaning, and
merely serves to create a node in the interior of
the wave function of the loosely bound state.

To obtain a second description of the projectile, that
differs from the first only in the interior,
we remove the spurious deeply bound state.
This eliminates the node in the wave function of
the physical ground state. However, we want to preserve
the ANC of the bound state as well as the phase shifts
obtained with $V_0$, because they play a major role
in breakup calculations \cite{TB04,TB05,CN06}.
As shown by Baye \cite{Bay87l,Bay87a}, this can be done using
a pair of supersymmetric transformations.
The first removes the deeply bound state,
without altering the other levels of the spectrum \cite{Suk85a}.
The second restores the phase shifts of $V_0$
that are modified by the first \cite{Bay87l,Bay87a}.
The resulting potential $V_2$ exhibits the same
spectrum as the initial $V_0$ but for the removed state.
It is also phase equivalent as it gives the same phase shifts.
Moreover, the ANC of the remaining bound states are identical
in both descriptions.

The supersymmetric elimination is performed only in
the partial wave of the removed state. Therefore,
$V_2$ depends strongly on the angular momentum even if
initially a single potential is chosen to describe all
partial waves. In the $lj$ partial wave, the new potential
reads \cite{Bay87a}
\beq
V_2^{lj}(r)=V_0^{lj}(r)-\frac{\hbar^2}{\mu}\frac{d^2}{dr^2}
\ln\int_0^r[u^0_{lj}(E_0,r')]^2dr',
\eeqn{2}
where $u^0_{lj}$ is the reduced radial wave function of the
removed state of energy $E_0$.
The potential $V_2$ differs from $V_0$ only in the
range of $u^0_{lj}$.
Similarly, the wave functions of the
Hamiltonian eigenstates are affected only at short distances.
After the supersymmetric elimination, the wave functions read \cite{Bay87a}
\beq
u^2_{lj}(E,r)=u^0_{lj}(E,r)-u^0_{lj}(E_0,r)
\frac{\int_0^r u^0_{lj}(E_0,r')u^0_{lj}(E,r')dr'}
{\int_0^r[u^0_{lj}(E_0,r')]^2dr'},
\eeqn{3}
where $u^0_{lj}$ and $u^2_{lj}$ are the reduced radial wave functions
obtained with $V_0$ and $V_2$, respectively,
for either bound or scattering states.
The asymptotic properties of $V_0$
(ANC of the remaining bound states, and phase shifts in the continuum)
are therefore preserved by these transformations.

\subsection{Descriptions of \ex{8}B and \ex{11}Be}\label{B8Be11}
Our description of \ex{8}B is based on the two-body model used in
Refs.~\cite{TNT01,MTT02}.
The parameters of the \ex{7}Be-$p$ potential are the same
as in those former papers but for the depth
of the central term in the $p3/2$ partial wave.
For that partial wave we consider a much deeper
Woods-Saxon form factor, which
reproduces not only the loosely bound state of \ex{8}B
but also a spurious deeply bound state (at $E_{0p3/2}=-57.2$~MeV).
This potential (including the Coulomb and centrifugal terms),
and the corresponding radial wave function of the physical ground state
are displayed in Figs.~\ref{f1}(a) and (c), respectively (dotted lines).

To obtain the second description of the nucleus,
we remove the spurious $p3/2$ state using the
supersymmetric transformations (see \Sec{susy}).
This gives us the new potential in the $p3/2$ partial wave
plotted as a full line in \fig{f1}(a).
The corresponding wave function of the physical
$p3/2$ state is shown in \fig{f1}(c) (full line).
As expected, both wave functions are identical above 4~fm,
where they coincide with their asymptotic behavior
(dashed line). However, they strongly differ in the interior,
where one exhibits a node, while the other is nodeless.

\bfig
\includegraphics[width=15cm]{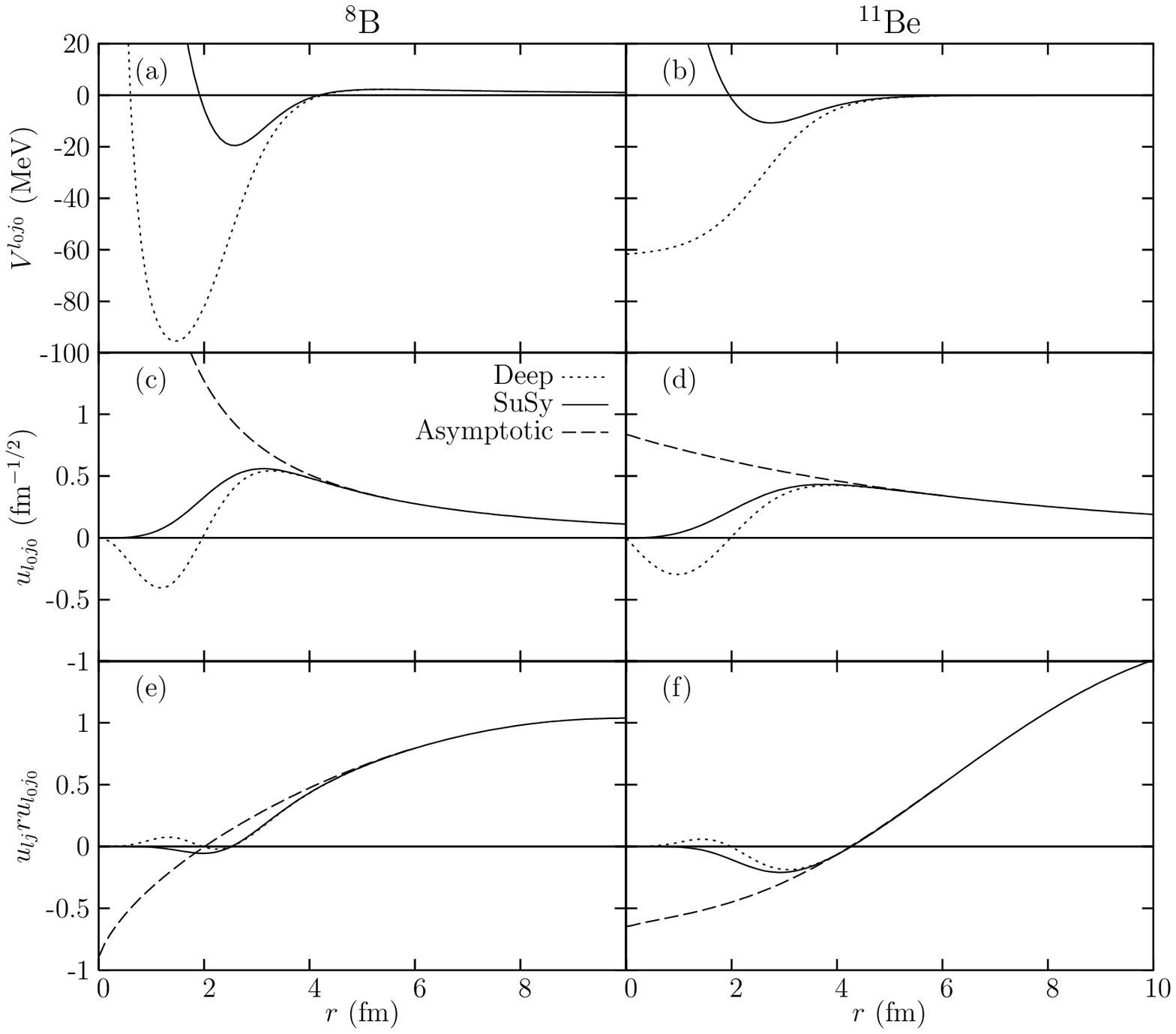}
\caption{Effective potentials describing the core-fragment interaction
for \ex{8}B (a), and \ex{11}Be (b).
Both the initial deep potentials (dotted lines), and their
supersymmetric partners (full lines) are depicted.
Radial wave functions of the loosely bound state of
\ex{8}B (c), and \ex{11}Be (d). Their asymptotic behavior
extrapolated down to $r=0$ is also shown (dashed line).
Radial integrand appearing in the
first-order approximation of the transition from
the ground state to a continuum state at $E=1$~MeV
in the $s1/2$ partial wave for \ex{8}B (e), and
in the $p3/2$ partial wave for \ex{11}Be (f) (see \Sec{fo}).
}\label{f1}
\efig

To describe \ex{11}Be, we consider the same \ex{10}Be-$n$ potential
as in \Ref{CGB04}. This model includes a Pauli forbidden
state in the $s1/2$ partial wave of the physical ground state
($E_{0s1/2}=-32.7$~MeV).
To obtain the second \ex{11}Be description, we simply
remove that state using the supersymmetric transformations.
The initial (dotted line) and supersymmetric (full line)
potentials are displayed in \fig{f1}(b).
The wave functions of the physical loosely bound state
are displayed in \fig{f1}(d) for both descriptions.
As in the \ex{8}B case, they differ at small distance, while
they concur above 4~fm.

\section{Breakup calculation}\label{results}

\subsection{Coulomb breakup at intermediate energy}\label{B8Pb}

Using the two descriptions of \ex{8}B,
we perform calculations of its breakup on lead at 44~MeV/nucleon.
It corresponds to the experiment performed at MSU by Davids
\etal \cite{Dav01}.
For this study, we use the dynamical eikonal approximation \cite{BCG05,GBC06}.
To simulate the interactions between the projectile components
and the lead target, we follow Mortimer \etal \cite{MTT02}.

The results of the calculations are illustrated in \fig{f2},
where the breakup cross section is plotted
as a function of the relative energy
between the \ex{7}Be core and the proton after dissociation.
The difference between the two calculations is completely
negligible (about 0.5~\%).
This confirms that breakup probes only the
tail of the projectile wave function,
since it is not sensitive to the large differences in
the interior.

\bfig
\includegraphics[width=10cm]{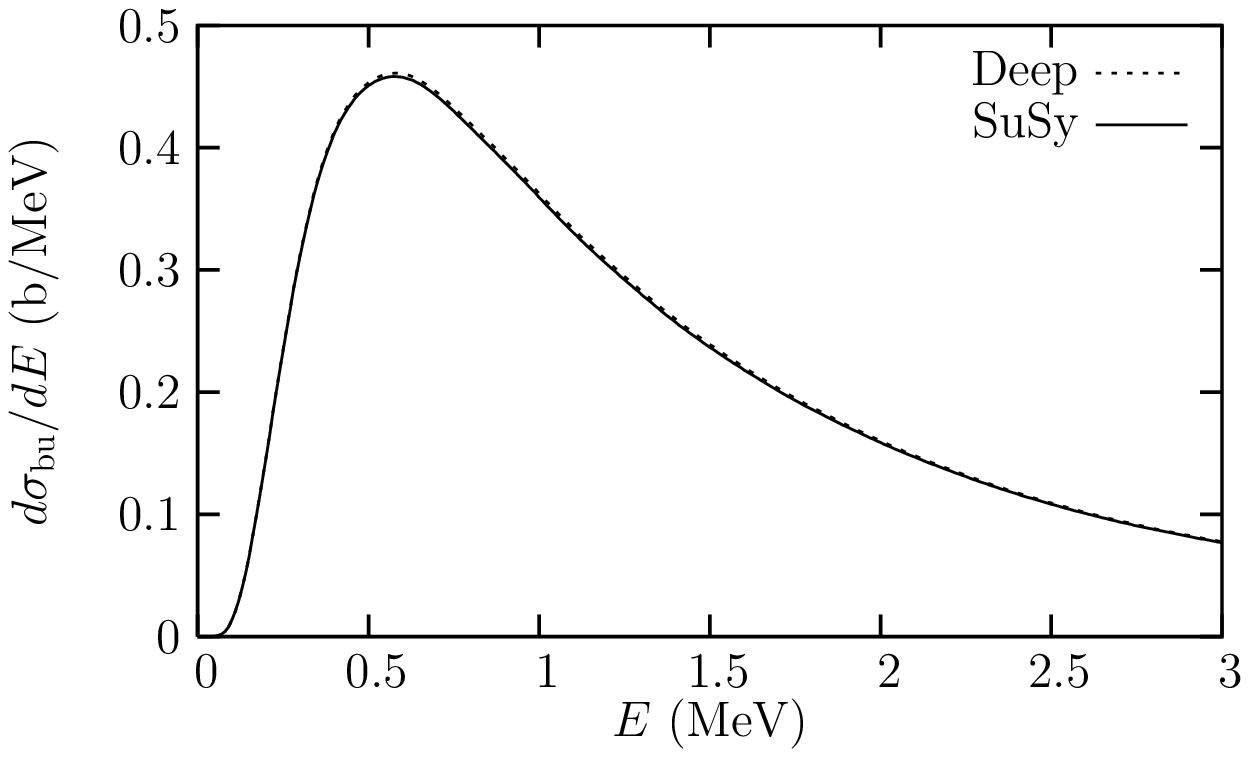}
\caption{
Breakup cross section of \ex{8}B on Pb at 44~MeV/nucleon
as a function of the energy between the proton and the \ex{7}Be
after dissociation.
The results are obtained using either the deep potential (dotted line)
or its supersymmetric partner (full line).
}\label{f2}
\efig

A partial wave analysis of this cross section shows that
the same result is obtained for the contribution of all partial
waves. The tiny difference observed in \fig{f2} is not due
to some accidental cancellation.
Negligible differences
are also observed in angular and parallel-momentum distributions.
Note that the presence of the spurious deeply bound state
in the $p3/2$ partial wave of the deep potential
does not affect the breakup calculation.
The cross section for the inelastic
process leading to the population of this state is indeed
of 8~$\mu$b, which is five orders of magnitude below the total
breakup cross section
(0.874~b for the deep potential, and 0.869~b for its supersymmetric partner).

The present result is in perfect agreement with
the analysis presented in \Ref{CBM03b}.
In that paper, it was found that eliminating the
forbidden bound states from the \ex{11}Be spectrum
has a very weak effect on the calculation of its breakup
on lead at 70~MeV/nucleon.

\subsection{First-order analysis}\label{fo}
With the aim of understanding qualitatively this result,
we make a first-order analysis.
In this approximation \cite{AW75}, the breakup
is assumed to occur in one step from the initial bound state
to the continuum. The reaction being Coulomb dominated, we
consider here only the Coulomb interaction,
which we expand into multipoles.
The breakup cross section can then be expressed analytically
as the sum of the contributions of the multipoles \cite{AW75}.
The same result is obtained within this framework: the difference
between both descriptions is negligible.

The dependence on the projectile description
of the contribution of the multipole $E\lambda$ comes from the radial integral
\beq
\mathcal{R}_{Elj}^{(\lambda)}=\int_0^\infty u_{lj}(E,r)
r^\lambda u_{l_0j_0}(E_0,r)dr,
\eeqn{4}
where $u_{lj}$ is the radial wave function describing
the $c$-$f$ continuum in the $lj$ partial wave at energy $E$,
and $u_{l_0j_0}$ is
the radial wave function of the initial bound state.
The integrand in \Eq{4} is displayed in \fig{f1}(e)
for the dominant $E1$ transition from the initial $p3/2$
bound state of \ex{8}B to the $s1/2$ continuum state computed at $E=1$~MeV.
This integrand for the $E1$ transition in \ex{11}Be
from the $s1/2$ bound state to the $p3/2$ continuum state at $E=1$~MeV
is depicted in \fig{f1}(f).
For both nuclei, the major contribution to $\mathcal{R}_{Elj}$ comes from
radii above 4~fm, where the wave functions
obtained with the deep and supersymmetric potentials coincide.
The interior of the wave functions contributes very little
to the integral mainly because of the large tail of the
wave function of the loosely bound state.
Moreover, the vanishing behavior of the radial wave functions
near the origin, multiplied by the $r^\lambda$ factor,
strongly reduces the contribution of the interior.
This explains qualitatively why the internal part of
the wave function has nearly no effect on breakup calculations.
It also indicates that one cannot choose any form factor
of the wave functions in the interior.
For example, one could use the extrapolation down to $r=0$
of the asymptotic behavior of the wave functions
[see dashed lines in Figs.~\ref{f1}(c) and (d)].
This leads to the dashed curves in Figs.~\ref{f1}(e) and (f).
In this case, the interior contribution is obviously overemphasized.
If non-vanishing wave functions are used, it is necessary
to resort to a cutoff at low radius, as Typel and Baur in \Ref{TB05}.

\subsection{Coulomb breakup at low energy}\label{B8Ni}
To see if the result of \Sec{B8Pb} holds at lower energy,
we calculate the breakup of \ex{8}B on nickel at 26~MeV.
It corresponds to the experiment performed at
Notre Dame \cite{Kol01}.
This reaction has been analyzed within the CDCC framework in
\Ref{TNT01}. For the present study, the calculations are
performed in a similar way using the code FRESCO \cite{FRESCO}.

\bfig
\includegraphics[width=10cm]{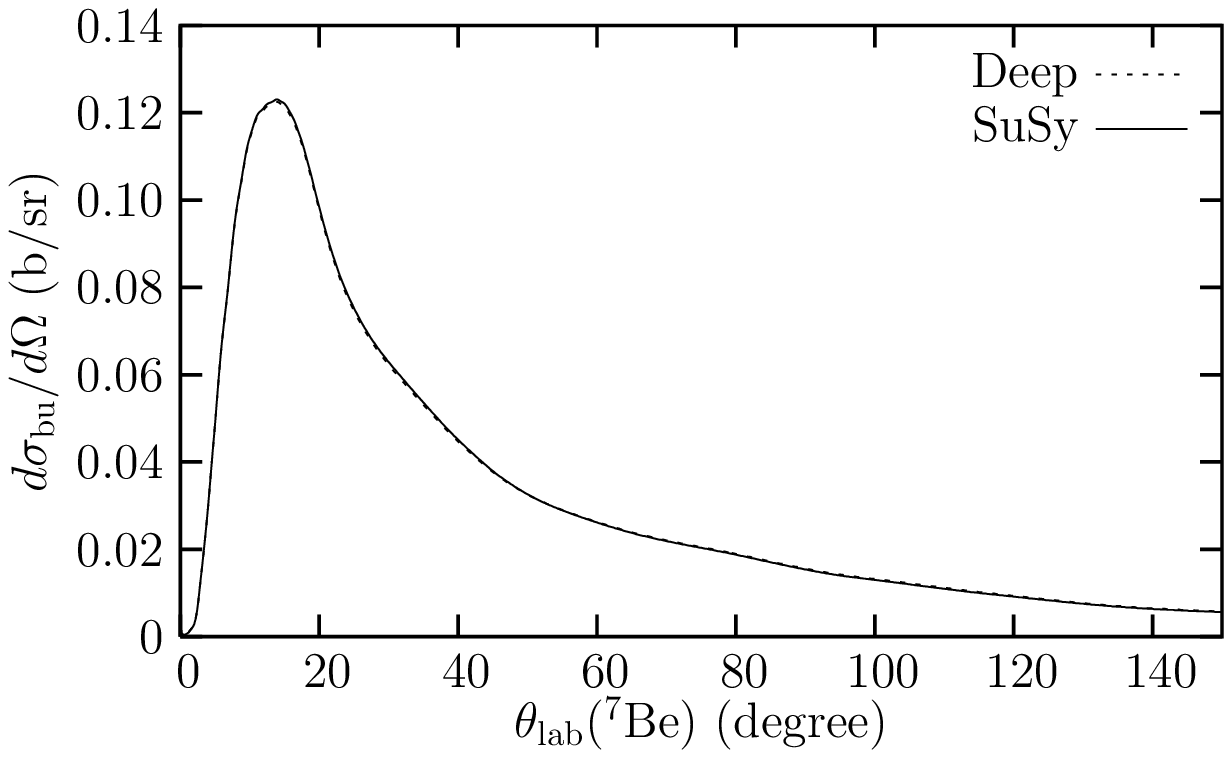}
\caption{
Breakup cross section of \ex{8}B on Ni at 26~MeV
as a function of the \ex{7}Be scattering angle after dissociation.
The results are obtained using either the deep potential (dotted line)
or its supersymmetric partner (full line).
The difference is so small that both curves are superimposed.
}\label{f3}
\efig

The angular distribution of the \ex{7}Be core
after breakup is shown in \fig{f3}.
The difference between both \ex{8}B descriptions is
even smaller than in the previous case: the two results are indistinguishable.
Interestingly this is true for the entire angular range.
In particular, even at large scattering angle,
where the nuclear interaction dominates, no difference
can be observed between both \ex{8}B descriptions.
The same conclusions are drawn from the energy distribution.

This additional result strengthens the conclusion of \Sec{B8Pb}
by extending its validity toward low energies.
The dissociation of loosely bound nuclei therefore never seems
to reflect the interior of the projectile.
Only the tail of the wave function is probed by such reactions.

\subsection{Nuclear-dominated breakup}\label{Be11C}
Finally, we investigate the influence of the interior
of the projectile wave function onto nuclear dominated
reactions. We choose the breakup of \ex{11}Be on carbon
at 68~MeV/nucleon. This reaction has been measured
at RIKEN by Fukuda \etal \cite{Fuk04}.
It has been successfully described using the dynamical
eikonal approximation \cite{CGB04,GBC06}, as well as
the CDCC technique \cite{HTA05}.
We use the former model with the same inputs as in \Ref{CGB04}.

The breakup cross section is plotted in \fig{f4} as a function
of the relative energy between the \ex{10}Be core and the neutron
after dissociation.
The sensitivity to the potential choice is similar
to the previous calculations. Both descriptions lead
to nearly identical cross sections. The same conclusion
can be drawn from angular and parallel-momentum distributions.
This result shows that even in nuclear-dominated breakup,
the calculations are not sensitive to the
internal part of the projectile wave function.
As in the previous cases, only the tail of the wave function
is probed.

\bfig
\includegraphics[width=10cm]{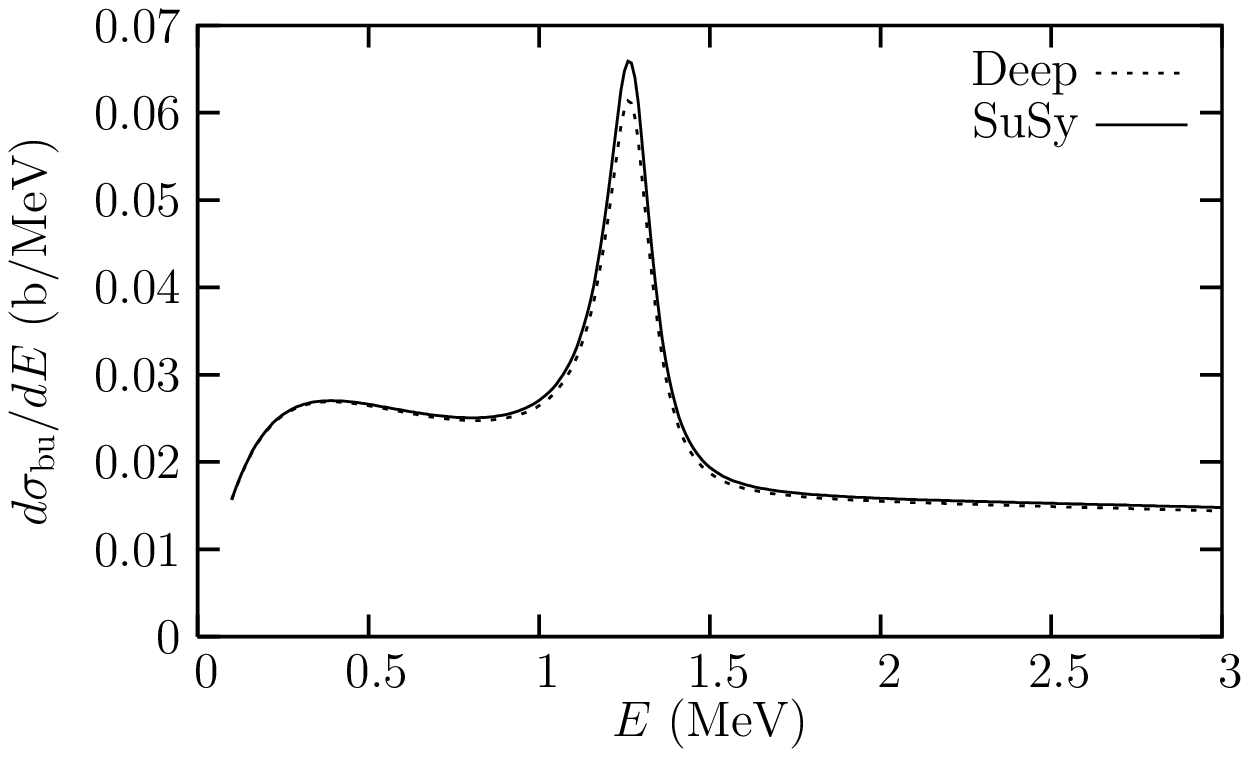}
\caption{
Breakup cross section of \ex{11}Be on C at 68~MeV/nucleon
as a function of the energy between the neutron and the \ex{10}Be
after dissociation.
The results are obtained using either the deep potential (dotted line)
or its supersymmetric partner (full line).
}\label{f4}
\efig

The only place where the two descriptions give slightly
different results is in the peak at 1.3~MeV, where they differ
by 8~\%. The peak is due to the presence of a $\fial^+$ resonance
at that energy \cite{CGB04}.
Since the difference appears only
in the $d5/2$ partial wave, where the resonance is reproduced,
we attribute this effect to the particular behavior of the
continuum wave function. This wave function
exhibits a huge peak near the origin
due to the resonant nature of the state.
In a first-order viewpoint, this peak increases
significantly the contribution of the interior to
the radial integral of \Eq{4}.
Consequently, a larger difference is observed.
This effect is in agreement with the analysis of \Ref{CBM03b},
where it has been observed that the excitation
to the $\half^-$ bound state of \ex{11}Be is more sensitive
than the breakup to the removal of the Pauli forbidden states.
In that case, both the initial and final states are bound.
As in the present case, the low-$r$ contribution is enhanced.
Therefore, due to the shorter range
of the final-state wave function,
the excitation to bound or resonant states
seems more sensitive to the interior of the wave function
than the non-resonant breakup.

\subsection{Comparison with projectile-target peripherality}\label{PT}
As mentioned in the introduction, other theoretical studies refer
to peripherality of reactions as insensitivity to small
projectile-target distances \cite{Bau84,SL98}. For compact
projectiles, no contributions from small impact parameters implies
the insensitivity to the internal part of the projectile wave
function. However, this may not be the case for loosely bound
projectiles and reaction processes with multi-step mechanisms.

Up to now, we have focused on peripherality relative to the
internal coordinate of the projectile. In this section, we analyse
the sensitivity of the breakup probability $P_{\rm bu}$ as a
function of the impact parameter $b$. In semiclassical models,
this breakup probability corresponds to the contribution of one
trajectory to the total breakup cross section. This is illustrated
in \fig{f5}.

\bfig
\includegraphics[width=10cm]{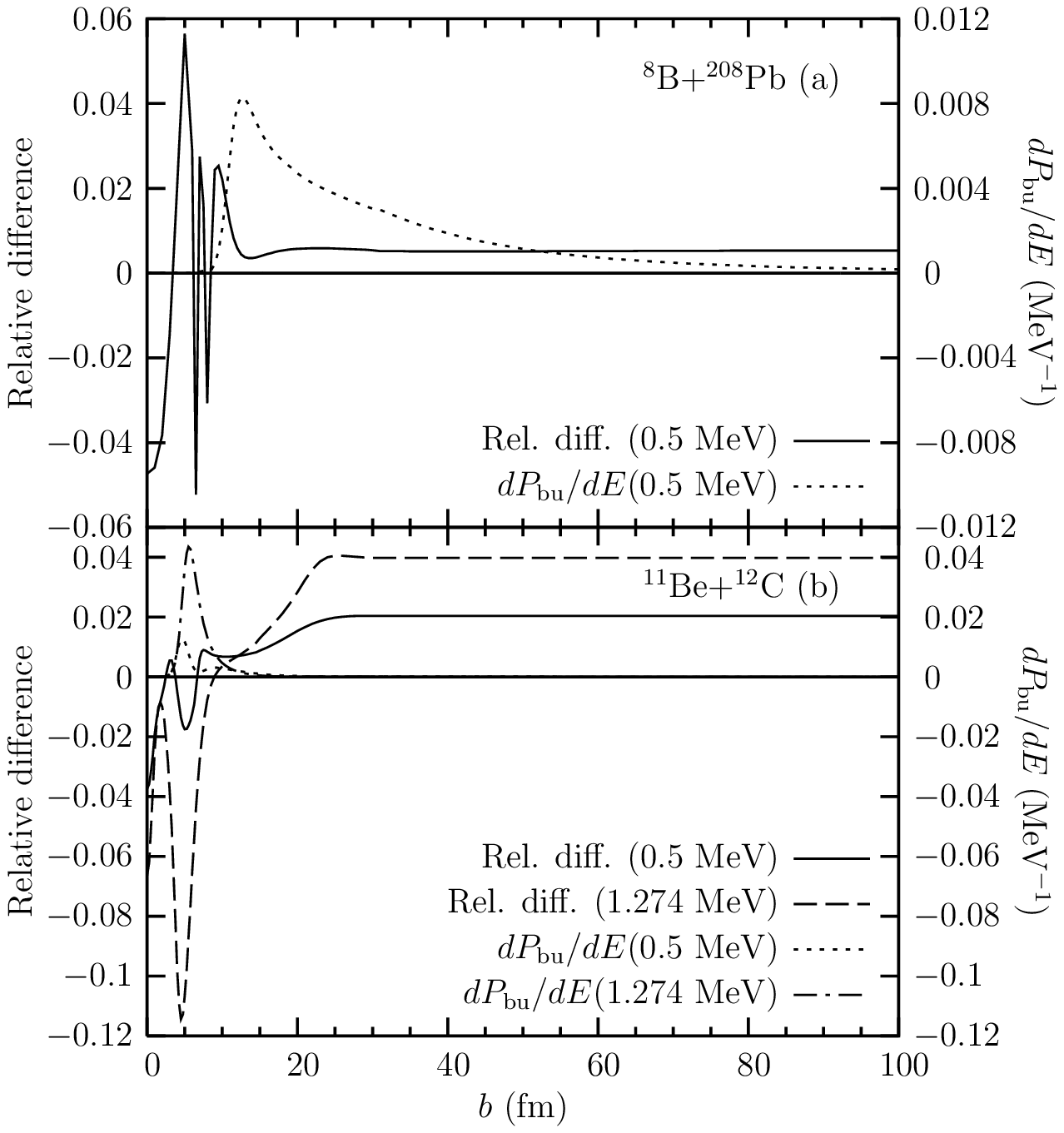}
\caption{
Relative difference between the deep and supersymmetric calculations
in the breakup probability as a function of the impact parameter.
The breakup probability obtained with the deep potential is shown as well.
(a) Breakup of \ex{8}B on Pb at 44~MeV/nucleon for $E=0.5$~MeV.
(b) Breakup of \ex{11}Be on C 68~MeV/nucleon for
$E=0.5$~MeV, and $E=1.274$~MeV (i.e. $\fial^+$-resonance energy).
}\label{f5}
\efig

The upper part (a) corresponds to the \ex{8}B breakup on lead at
44~MeV/nucleon for a relative energy $E=0.5$~MeV between the
\ex{7}Be core and the proton after dissociation. The breakup
probability computed using the deep potential is displayed in
dotted line. The full line corresponds to the relative difference
between the results obtained with the deep potential and its
supersymmetric partner. The first important point to notice is
that the reaction occurs at very large impact parameters: most of
the contributions occur beyond impact parameter of 15 fm.
Secondly, in the region where there are contributions to the cross
section, there is less than 1 \% difference between the
calculations using the deep projectile potential and the
supersymmetric partner, confirming no sensitivity to the
projectile wave function in the interior. This is a case where
the reaction is peripheral relative to both, projectile-target and
core-fragment distances.
However, the two peripheralities seem disconnected.
Beyond 15~fm, the relative difference between the
deep and supersymmetric potentials is indeed very constant.
It does not decrease with $b$: the sensitivity to the interior of the
projectile description is the same at large projectile-target
distances ($b\sim100$~fm) than at intermediate ones ($b\sim15$~fm).

We perform the same analysis for the dissociation of \ex{11}Be on
carbon at 68~MeV/nucleon. The breakup probability, and relative
difference between the two descriptions are displayed
in \fig{f5}(b) for two relative energies $E$:
0.5~MeV, and 1.274~MeV, the energy of the $\fial^+$
resonance. In this case,
the dominant contribution to the breakup cross section comes from
impact parameters around $b=5$~fm.
Expectedly, this nuclear dominated dissociation is less peripheral
than Coulomb breakup.
However, the difference between both descriptions remains
negligible, except for the resonance energy, as seen in
\Sec{Be11C}.
Therefore, even though the reaction appears to be surface peaked in
the projectile-target distance, it is still insensitive to the
description of the projectile at small core-fragment distances.

\section{Conclusion}\label{conclusion}
In this paper, the sensitivity of elastic breakup
to the internal part of the projectile wave function is analyzed.
We compare calculations of dissociation of loosely bound nuclei
performed with two descriptions of the projectile,
which differ only in the interior of the wave function.
These descriptions are obtained from phase equivalent
potentials constructed with supersymmetric
transformations \cite{Bay87l,Bay87a}.

We start from an initial deep potential adjusted
to reproduce, not only the loosely bound state of the
projectile, but also a spurious deeply bound state.
Due to the presence of that unphysical state,
the projectile wave function exhibits a node near the origin.
A second description is obtained by removing
the spurious state. The wave function therefore
no longer exhibits a node.
This removal is achieved by supersymmetric
transformations, which preserve the asymptotic
properties of the initial potential \cite{Bay87l,Bay87a}
(i.e. ANC and phase shifts).
This enables us to analyze the sensitivity of the calculations
only to the interior of the wave function.

We study this effect in the dissociation
of two halo nuclei: \ex{8}B, and \ex{11}Be.
Both Coulomb and nuclear induced reactions are
considered, as well as intermediate and low energy regimes.
In all the analyzed cases, no difference
between the deep potential and its supersymmetric partner
is obtained, although the corresponding wave functions
of the projectile strongly differ in the interior
(presence or absence of a node).
This is observed in several observables such as
energy, angular, and parallel-momentum distributions.
We conclude that the elastic breakup
of loosely bound nuclei is peripheral in the sense
that it is not sensitive to
the interior of the description of the projectile.
Only the asymptotic properties of the
projectile model are probed through these reactions;
namely the ANC of the bound state,
as suggested in Refs.~\cite{TCG01,TCG04},
and the phase shifts in the continuum \cite{TB04,TB05,CN06}.
Therefore, two descriptions of the projectile corresponding to the same ANC
but different spectroscopic factors will lead to the same theoretical
prediction. This shows that spectroscopic factors extracted from breakup
measurements will contain an error due to the imprecision of the projectile
description at short distances, which is not probed by breakup reactions.
The extraction of an ANC from breakup is less model dependent.
Nevertheless, it will still contain uncertainties
associated with the projectile-target interactions and the description of
the projectile continuum.

\begin{acknowledgments}
We thank J.~A.~Tostevin for interesting discussions on this topic.
This work has been initiated
during the program INT-05-3 organized by the Institute for Nuclear Theory
at the University of Washington. P.~C. thanks this institute
for its hospitality and the Department of Energy for partial
support during the program.
He also acknowledges the support of the Natural Sciences
and Engineering Research Council of Canada (NSERC). Support
from the National Superconducting Cyclotron Laboratory (NSCL)
at Michigan State University and from the
National Science Foundation grant PHY-0456656 is
acknowledged.
\end{acknowledgments}


\end{document}